\documentstyle[12pt]{article}
\makeatletter                                       %
\@addtoreset{equation}{section}                     % Style of
\makeatother                                        %
\begin{document}
\begin{flushright}
hep-th/9511001 \\
October 1995 \\
\end{flushright}
\vspace{25 pt }
\begin{center}
{\Large\bf
    Quantization of 2D Abelian Gauge Theory\\
   without the Kinetic Term of Gauge Field \\
   as Anomalous Gauge Theory \\
}
\end{center}
\vfill
\def\thefootnote{\alph{footnote}}
\begin{center}{\sc M.Koseki}
\footnote{E-mail address: koseki@muse.hep.sc.niigata-u.ac.jp}
,{\sc R.Kuriki}
\footnote{E-mail address: r-kuriki@phys.titech.ac.jp}\\
\vspace{10 pt}
{\em Graduate School of Science and Technology, Niigata University,
Niigata, 950-21, Japan}
{\em Department of Physics, Tokyo Institute of Technology, Tokyo 152, Japan}
\end{center}
\vfill
%%%%%%%%%%%%%%%%%%%%%%%%%%%%%%%%%%%%%%%%%%%%%%%%%%%%%%%%%%%%%%%%%%%%%%%%%%%%%%%

\abstract

The massless Schwinger model
without the kinetic term of gauge field has gauge anomaly.
We quantize the model as an anomalous gauge theory in the most general class
of gauge conditions.
We show that the gauge field becomes a dynamical variable because of
gauge anomaly.

\def\thefootnote{\fnsymbol{footnote}}
\vfill
\newpage
%%%%%%%%%%%%%%%%%%%%%%%%%%%%%%%%%%%%%%%%%%%%%%%%%%%%%%%%%%%%%%%%%%%%%%%%%%%%%%%
%
%     SYMBOLS ARE DEFINED BELOW
%
%%%%%%%%%%%%%%%%%%%%%%%%%%%%%%%%%%%%%%%%%%%%%%%%%%%%%%%%%%%%%%%%%%%%%%%%%%%%%%%
%
%   Greek letters
%
\def\a{\alpha}     \def\b{\beta}         \def\g{\gamma}   \def\d{\delta}
\def\e{\epsilon}   \def\ve{\varepsilon}  \def\z{\zeta}    \def\c{\theta}
\def\vc{\vartheta} \def\i{\iota}         \def\k{\kappa}   \def\l{\lambda}
\def\m{\mu}        \def\n{\nu}           \def\x{\xi}      \def\p{\pi}
\def\r{\rho}       \def\s{\sigma}        \def\t{\tau}     \def\f{\phi}
\def\vp{\varphi}   \def\v{\psi}          \def\w{\omega}   \def\h{\eta}

\def\G{\Gamma}     \def\D{\Delta}        \def\C{\Theta}   \def\L{\Lambda}
\def\X{\Xi}        \def\P{\Pi}           \def\SS{\Sigma}  \def\F{\Phi}
\def\V{\Psi}       \def\W{\Omega}        \def\U{\Upsilon}
%
%   mathematical symbols
%
\def\dl{\partial}   \def\7{\bigtriangledown}    \def\3{\bigtriangleup}
%
%%%%%%%%%%%%%%%%%%%%%%%%%%%%%%%%%%%%%%%%%%%%%%%%%%%%%%%%%%%%%%%%%%%%%%%%%%%%%%%
%
%n     OTHER USEFUL DEFINITION
%
%%%%%%%%%%%%%%%%%%%%%%%%%%%%%%%%%%%%%%%%%%%%%%%%%%%%%%%%%%%%%%%%%%%%%%%%%%%%%%%
\newcommand{\bi}{\bibitem}
\newcommand{\be}{\begin{eqnarray}}
\newcommand{\ee}{\end{eqnarray}}
\newcommand{\nn}{\nonumber}
\newcommand{\ds}{\displaystyle}
%%%%%%%%%%%%%%%%%%%%%%%%%%%%%%%%%%%%%%%%%%%%%%%%%%%%%%%%%%%%%%%%%%%%%%%%%%%%%%%

\section{Introduction}

\pagestyle{plain}
\setcounter{equation}{0}

Two-dimensional field theories coupled to gauge fields
without kinetic terms have recently been investigated
in several context [1-4].
There are two motivations to study such gauge theory.

One of them is the expectation
that the models may provide realizations of the GKO coset construction
of conformal field theories\cite{GKO}. The Schwinger model without
kinetic term of gauge field corresponds to the $U(1)/U(1)$ coset model
in the language of GKO coset construction.
Actually, Itoi and Mukaida studied the model \cite{IM}, and
they asserted that the model was a topological field theory \cite{TO1}.

Another motivation to study the gauge theory without a kinetic term is
the expectation that the analysis of such gauge theory will help us to
investigate non-critical string theories.
Actually, Polyakov studied such a gauge theory as a toy model of
two-dimensional gravity \cite{NK2}.

The symmetries in the Schwinger model without kinetic term of
gauge field  is completely different from one in the usual Schwinger model.
Moreover, the Schwinger model without a kinetic term of gauge field has
gauge anomaly, but there is no gauge anomaly in the usual Schwinger
model \cite{sch}.
Therefore, one can treat this model as an anomalous gauge theory.

There are two purposes of this paper.
The one of them is to formulate the Schwinger model without a kinetic
term of gauge field as an anomalous gauge theory in the most general
class of gauge fixing conditions.
The other is to show that a degree of freedom of gauge field becomes a
dynamical variable because of gauge anomaly, though one can eliminate the
all degree of freedom of gauge field by local symmetries at classical
level.

The strategy to achieve our aim is as follows.
To avoid the confusion between this model and the usual Schwinger model,
we show that gauge anomaly exists in this model in  Appendix A.
In Section 2, the Hamiltonian formalism developed by Batalin and Fradkin
( BF formalism ) is applied to convert the constraints effectively from
second class to first class. A generally BRST
quantization scheme by the extended phase space of Batalin, Fradkin and
Vilkovisky (BFV) is constructed.
In Section 3, we investigate the theory under two types of gauge fixing
conditions.
In the light-cone gauge, we show that the degree of freedom of gauge field
becomes a
dynamical variable at quantum level.
Conclusion are given in Section 4.

%%%%%%%%%%%%%%%%%%%%%%%%%%%%%%%%%%%%%%%%%%%%%%%%%%%%%%%%%%%%%%%%%%%%%%%%%%%%%%%

\section{Symmetrization and BRST invariant effective action in the most
general class of gauges}
In this section,
first, we investigate properties of constraints in this model.
The constraints belong to the first class at the classical level,
but these constraints belong to the second class at
quantum level because of gauge anomaly.
We convert them into first class by introducing an extra degree of
freedom.
Finally, the converted system is quantized in the extended phase space,
and a BRST invariant effective action is constructed.

\vspace{0.3cm}

First, we summarize the properties of the symmetry in the Schwinger
model without the kinetic term of gauge field.
The Lagrangian density of the Schwinger model without the kinetic term
of gauge field is given by
\be
\begin{array}{rl}
{\cal L}&=\bar{\psi}\gamma^{\mu}\left({i}\partial_{\mu}+A_{\mu}\right)\psi \\
        &=\psi^{\dag}_+\left({i}\partial_{-}+A_{-}\right)\psi_{+}
         +\psi^{\dag}_{-}\left({i}\partial_{+}+A_+\right)\psi_{-}, \\
\end{array}\label{action}
\ee
where $\psi$ and $A_{\mu}~\left(\mu=0,1\right)$ are Dirac field and
$U\left(1\right)$ gauge field
\footnote{We use the following conventions in two dimensions;
\[{\gamma}^0={\sigma}^1,{\gamma}^1=i{\sigma}^2,
{\gamma}^5={\gamma}^0{\gamma}^1,\partial_{\pm}=\partial_0\pm\partial_1,
A_{\pm}=A_0\pm A_1\]
\[\dot{F}=\partial_0 F,~~F{'}=\partial_{1}
F,~~\psi^{\dag}=\left(\psi^{\dag}_+\psi^{\dag}_-\right)\]
 and 2 dimensional coordinates $x^{\mu}\left(\mu=0,1\right)$
are denoted by $\left(t,x\right)$ and take
$-\infty< x <\infty,\quad -\infty< t <\infty $.}, respectively.
This Lagrangian density has $U(1)_V\times U(1)_A$ local symmetries at
classical level\cite{YW}.
Although the model has two local symmetries at classical level, no
regularization of it can preserve both the symmetries simultaneously.
Not all of these symmetries can survive because of anomalies in ordinary
configuration space.

The equal time super-Poisson bracket among these variables
are given by
\be
\begin{array}{r l}
\left\{A_{\pm}\left(x,t\right),\pi^A_{\pm}\left(y,t\right)\right\}=
 &\delta\left(x-y\right),\\
\left\{\psi_{\pm}\left(x,t\right),\psi^{\dag}_{\pm}\left(y,t\right)\right\}=
&-i\delta\left(x-y\right).
\end{array}
\ee
In the canonical description, the local symmetries manifest themselves as
first class constraints according to Dirac 's  classification \cite{Dirac}.
Denoting the canonical momenta for $A_\pm$ by $\pi^A_\pm$,
we write the primary constraints as
\be
 \pi^A_\pm\approx{0},\label{primary}
\ee
where $\pi^A_\pm\equiv{\partial{\cal L}}{/}{\partial\dot{A}_\pm}$.
Consistency condition for the primary constraints leads the secondary
constraints,
\be
{\dot { \pi^A_\pm }} = -\left\{ H_0 ,\pi^A_\pm \right\} = j_\pm\approx{0},
\label{secondary}\ee
where $j_\pm \equiv\psi^{\dag}_\pm\psi_\pm $. Here the Hamiltonian $H_0$ is
\be
 H_0=\int dy \left[ i\left(\psi^{\dag}_+\psi_+{'}-\psi^{\dag}_+\psi_-{'}\right)
     -A_{-}j_{+}-A_{+}j_{-}+\lambda_+\pi^A_{+}+\lambda_{-}\pi^A_{-} \right],
\label{hamiltonian}
\ee
where $\lambda_\pm$ are Lagrange multiplier for the primary constraints.
The consistency condition for $j_\pm$ is automatically satisfied and all
super-Poisson brackets vanish.
Therefore all constraints are first-class at classical level.
The canonical theory of this system is characterized by the first-class
constraints
\footnote{The variable $\psi^{\dag}_\pm$ is taken as the canonical momentum
conjugate to $\psi_\pm$. } (\ref{primary}) (\ref{secondary}).

For the quantization of the model
\footnote{We use $\hbar=1$ and omit the normal ordering symbol
as given in Appendix A. For definitions of the equal-time anti-commutation
relations, we simply replace the super-Poisson bracket $\left\{A,B\right\}$ by
the graded commutator $-i\left[A,B\right]$ defined as
$\left[A,B\right]=AB-BA\left(-\right)^{\epsilon\left(A\right)
\epsilon\left(B\right)}$. Where $\epsilon\left(A\right)$ is the grassmannian
parity of $A$. This naive prescription can be justified
in the BF formalism, because we do not use the Dirac bracket.}
, we replace the super-Poisson bracket by the super-commutator.
To regularizing operator products, we adopt the
normal ordering prescription. The definition of
normal ordering in Appendix B is described.
Now the equal-time commutators between operator $j_\pm$ does not vanish,
\be
\left[j_\pm\left(x,t\right),j_\pm\left(y,t\right)\right]
 =\pm\frac{i}{2\pi}\partial_x\delta\left(x-y\right).\label{schwinger}
\ee
The term in r.h.s. is so-called Schwinger term which represents the
commutator anomaly of gauge algebra.
The form of this term agrees with a form of the cohomology given in
Appendix A.
Namely, the form of the Schwinger term is independent of the choice of
regularizations and gauge-fixing, except for the overall coefficient
$\pm{i}/2\pi$.
The consistency condition for $j_{\pm}$  is given by
\be
 \left[j_{\pm}\left(x,t\right),H_0\right]
 =\pm i{j}_\pm{'}\left(x,t\right)\mp\frac{i}{2\pi}A_\mp{'}\left(x,t\right).
\label{new}
\ee
To guarantee the consistency condition for $j_\pm$, new constraints are
imposed as follows
\be
A_\pm{'}\left(x,t\right)\approx{0}.
\label{neww}\ee
Moreover the consistency condition gives ``$\lambda_\pm=0$''.
Because of the anomaly, these constraints become second class all together.

\vspace{0.5cm}

Next, we modify the theory to recover all the classical local symmetries
violated by anomalies.
We apply the BF formalism to recover the classical local symmetries.
This can be carried out without affecting the physical contents of the
original theory, by introducing extra degrees of freedom which can be
gauged away by the recovered local symmetry.
Following the general idea of BF\cite{BF}, we introduce a canonical
pair of bosonic fields $\left(\pi_\theta,\theta\right)$, which is called
as BF fields henceforth.
They are assumed to satisfy the following commutation relation
\be
 \left[ \theta\left(x,t\right),\pi_\theta\left(y,t\right)\right]
  =-i\delta\left(x-y\right).
\ee
Notice that the BF field must have negative metric in this model.
The constraint can be modified by adding to $j_\pm$ an appropriate term
containing BF field to cancel the Schwinger term in Eq.(\ref{schwinger}).
The new modified constraint is given by
\be
 \tilde{j}_\pm\equiv{j}_\pm+k\left(\pi_{\theta}\pm\theta{'}\right)\approx{0}.
\label{jchil}\ee
if $k^2={1}{/}{4\pi}$\footnote{If the canonical commutation relation
$\left[\theta\left(x,t\right),
\pi_\theta\left(y,t\right)\right]=i\delta\left(x-y\right)$ is imposed,
we have
$k^2=-{1}{/}{4\pi}$, with $k$ being imaginary.
But $k$ must be real because of Hermiticity of Lagrangian density.},
then the algebras of modified constraints become
\be
\begin{array}{rl}
\left[\tilde{j_\pm}\left( x,t\right) ,\tilde{j_\pm}\left( y,t\right)\right ]
  =&0.\\
\end{array}
\ee
Cancelation of the second term in r.h.s.of Eq.(\ref{new}) also occurs by
a modification of the Hamiltonian.
The modified Hamiltonian is constructed by adding some polynomials of
the BF fields to the Hamiltonian in (\ref{hamiltonian}).
Our choice is expressed by
\be
 \tilde{H}_0=H_0
   -\int dx\bigl[\frac{1}{4}\left(\pi_\theta+\theta{'}\right)^2
   +\frac{1}{4}\left(\pi_\theta-\theta{'}\right)^2\bigr].
\label{Hchil}\ee
The  Eq.(\ref{jchil}) and Eq.(\ref{Hchil}) become the first class
\be
\begin{array}{rl}
\left[\tilde{j_\pm}\left( x,t \right) ,\tilde{H}_0 \right]
  =&\pm i\tilde{j_\pm}^\prime \left( x,t \right) .\\
\end{array}\label{qa}
\ee
Here, the consistency condition of $\tilde{j}_\pm$ is satisfied without
imposing new constraints in Eq.(\ref{neww}).
Therefore owing to the extra degree of freedom $\pi_\theta$ and
$\theta$, the original second class system becomes an effectively first
class system.

\vspace{0.3cm}

Then, according to the generalized Hamiltonian formalism developed by
Batalin, Fradkin and Vilkovisky (BFV formalism) \cite{BFV}\footnote{For
review, see\cite{MHBF}}, we introduce canonical sets of ghosts and
anti-ghosts along with the auxiliary fields, for each constraint
$\pi^A_\pm\approx{0}$ and $\tilde{j}_\pm\approx{0}$.
In other word, we define the extended phase space (EPS),
\be
\begin{array}{c}
\tilde{j}_{\pm} : \left({\cal C}^{\pm},\bar{{\cal P}}_{\pm}\right),
\left({\cal P}^{\pm},\bar{{\cal C}}_{\pm}\right),
\left(N^{\pm}, B_{\pm}\right);\nonumber\\
{\pi}^{A}_{\pm} : \left({\cal C}^{\pm}_{\pi},
\bar{{\cal P}}^{\pi}_{\pm}\right),\left({\cal P}^{\pm}_{\pi},
\bar{{\cal C}}^{\pi}_{\pm}\right),\left(N^{\pm}_{\pi},B^{\pi}_{\pm}\right).\\
\end{array}\label{EPS}
\ee
The off-shell nilpotent BRST charge is given by
\be
\begin{array}{crcl}
 &Q\; &=&Q_-+Q_+ ; \\
 &Q_\pm &\equiv
  &\displaystyle\int dx \left[{\cal C}^{\pm}\tilde{j}_{\pm}
   +{\cal C}^{\pm}_{\pi}\pi^A_{\pm}+{\cal P}^{\pm}B_{\pm}
    +{\cal P}^{\pm}_{\pi}B^{\pi}_{\pm}\right]\\
\end{array}\label{brst}
\ee
This charge $Q$ generates the BRST transformation of the fundamental variables.
\be
\begin{array}{rlrlrlrl}
\delta\psi_\pm=&-i{\cal C}^{\pm}\psi_{\pm},\quad
 &\delta\psi^{\dag}_{\pm}=&i{\cal C}^{\pm}\psi^{\dag}_{\pm},\quad
 &\delta A_{\pm}=&{\cal C}^\pm_\pi,\quad &\delta{\cal C}^\pm_\pi=&0,\\
\delta{N}^\pm=&{\cal P}^\pm,\quad &\delta{\cal P}^\pm=&0,\quad
 &\delta{N}^\pm_\pi=&{\cal P}^\pm_\pi,\quad &\delta{\cal P}^\pm_\pi=&0,\\
\delta\bar{{\cal C}}_\pm=&-B_\pm,\quad &\delta{B}_\pm=&0,\quad
 &\delta\bar{{\cal C}}^\pi_\pm=&-B^\pi_\pm,\quad &\delta{B}^\pi_\pm=&0,\\
\delta\bar{{\cal P}}_\pm=&-{i}\tilde{j}_\pm,&\delta{\cal C}^\pm=&0,\quad
 &\delta\bar{{\cal P}}^\pi_\pm=&-\pi^A_\pm\quad &\delta\pi^A_\pm=&0,\\
\end{array}
\ee
\vspace{-22 pt}
\be
\begin{array}{rlrl}
\delta\theta=&-k\left({\cal C}^++{\cal C}^-\right),\quad
 &\delta\pi_{\theta}=&-k\left({\cal C}^+-{\cal C}^-\right){'},\hspace{90pt}\\
\end{array}\nn
\label{brsttr}
\ee
and the BRST invariant Hamiltonian, which is called as the minimal
Hamiltonian,
\be
H_{min}=\tilde{H}_0+\int dx\left(\overline{\cal P}_+{\cal C}^+{'}-
\overline{\cal P}_{-}{\cal C}^{-}{'}\right).\label{Hmin}
\ee
The second term in (\ref{Hmin}) is required for the minimal Hamiltonian
to respect BRST invariance.

The dynamics of the system are controlled by the BRST invariant total
Hamiltonian $H_T$, which consists of the minimal Hamiltonian and the
gauge-fixing term.
In the present case, $H_T$ is given by
\be
H_T=H_{min}+\frac{1}{i}\left[Q,\Psi\right].
\ee
where $\Psi$ is a gauge fermion. The standard form of
$\Psi$ is given by
\be
\begin{array}{rrcl}
 &\Psi&=&\Psi_++\Psi_- ;\\
 &\Psi_\pm &\equiv
   &\displaystyle \int dx\left[ \overline{\cal C}_{\pm}\chi^{\pm}
    +\overline{\cal C}^{\pi}_{\pm}\chi^{\pm}_{\pi}
     +\overline{\cal P}_{\pm}N^{\pm}
       +\overline{\cal P}^{\pi}_{\pm}N^{\pm}_{\pi}\right], \\
\end{array}\label{gaugefermion}
\ee
where $\chi^\pm,\chi^\pm_\pi$ denote the gauge conditions imposed on
dynamical variables.
\footnote{
To construct the effective action, we choose the standard form of
gauge fermion shifted by \\
$\Psi_\pm\rightarrow\Psi_\pm +\int dx\left(\overline{\cal C}_{\pm}\dot{N}^{\pm}
+\overline{\cal C}^{\pi}_{\pm}\dot{N}^{\pm}_{\pi}\right)$.

This just cancels the Legendre terms $\int dx\left[\overline{\cal C}_{\pm}
\dot{\cal P}^{\pm}+\overline{\cal C}^{\pi}_{\pm}\dot{\cal P}^{\pm}_{\pi}
+B_{\pm}\dot{N}^{\pm}+B^{\pi}_{\pm}\dot{N}^{\pm}_{\pi}\right]$
in the constructed effective action.
}

The BRST invariant effective action can be obtained as
\be
\begin{array}{rcl@{}l}
S_{eff}&=&\displaystyle \int d^2x &\left[
   i\psi^{\dag}_{\pm}\dot{\psi}_{\pm}+\pi^A_{\pm}\dot{A}_{\pm}
   +\bar{{\cal P}}_{\pm}\dot{{\cal C}}^{\pm}_{\pi}
   +\bar{{\cal P}}^{\pi}_{\pm}\dot{{\cal C}}^{\pm}_{\pi}
   -\pi_\theta\dot{\theta}\right] -\displaystyle \int{dt H_T}\\
&=&\displaystyle \int dxdt & \left[
   {\psi}^{\dag}_{+}\left({i}\partial_{-}-N^{+}\right)\psi_{+}
   +{\psi}^{\dag}_{-}\left({i}\partial_{+}-N^{-}\right)\psi_{-} \right.\\
&&&-\pi_{\theta}\dot{\theta}
   +\frac{1}{2}\left(\pi^2_\theta+{\theta{'}^2}\right)
   -k\left(\pi_\theta+\theta{'}\right)N^+
   -k\left(\pi_\theta-\theta{'}\right)N^-\\
&&&+\pi^A_+\left(\dot{A}_+-N^+_{\pi}\right)
   +\pi^A_-\left(\dot{A}_--N^-_\pi\right)\\
&&&+\bar{{\cal  P}}_+\left(\partial_{-}{\cal C}^+-{\cal P}^+\right)
   +\bar{{\cal P}}_{-}\left(\partial_{+}{\cal C}^{-}-{\cal P}^-\right)
   +\bar{{\cal P}}^{\pm}_{\pi}\left(\dot{{\cal C}}^{\pm}_{\pi}
   -{\cal P}^{\pm}_{\pi}\right) \qquad\\
&&&-\biggl.\left(B_{\pm}\chi^{\pm}+B^{\pi}_{\pm}\chi^{\pi}_{\pm}
    +\bar{{\cal C}}^{\pi}_{\pm}\delta\chi^{\pi}_{\pm}
     +\bar{{\cal C}}_{\pm}\delta\chi^{\pm}\right) \biggr].\\
\end{array}\label{eff}
\ee
Since the auxiliary fields $N^\pm$  play the roles of
$-A_\mp$ in Eq.(\ref{eff}), respectively,
The following gauge conditions are imposed,
\begin{eqnarray}
\chi^\pi_\pm={N}^\pm+{A}_\mp.\label{geom}
\end{eqnarray}
This gauge fixing condition is essential to geometrize the gauge fields.
As for the rest of the gauge conditions $\chi_\pm$,
we assume that they and their BRST transformations are independent of,
$\pi^A_\pm, \pi_\theta,B^\pi_\pm,\bar{{\cal P}}_\pm,\bar{{\cal P}}^\pi_\pm,
\bar{{\cal C}}^\pi_\pm$.
By taking the variation of Eq.(\ref{eff}) with respect to variables,
we can derive equations of motion for $\pi^A_\pm,\bar{{\cal P}}_\pm,
\bar{{\cal P}}^\pi_\pm,B^\pi_\pm,\bar{{\cal C}}^\pi_\pm$ and $\pi_\theta$,
\be
\begin{array}{rl}
&\dot{A}_\pm=N^\pm_\pi,\quad\partial_\mp{\cal C}^\pm={\cal P}^\pm,\quad
  \dot{{\cal C}}^\pm_\pi={\cal P}^\pm_\pi,\quad{N}^\pm=-A_\mp,\\
&{\cal P}^\pm=-{\cal C}^\mp_\pi,\quad
  \dot{\theta}-\pi_\theta-{k}\left(A_-+A_+\right)=0.\\
\end{array}\label{eqmotion}
\ee
The BRST transformations in the configuration space is given by replacing
the parameters of infinitesimal gauge transformation with ghost fields.
There exist two independent ghost variables
$C_A,C_V$ in the configuration space, which correspond to $U(1)_A$
symmetry and the $U(1)_V$ symmetry respectively.
After imposing the gauge condition in Eq.(\ref{geom}), we obtain the
geometrization for the gauge field and the ghost variables in EPS as
\be
\partial_{\mp}{\cal C}^\pm=-{\cal C}^\mp_\pi,
\ee
And the BRST transformation for $A_\pm$ in the EPS is given by
\be
\delta{A}_\pm=-\partial_\pm{\cal C}^\mp.
\ee
The BRST transformations for $\psi_\pm$ and $A_\pm$ in configuration
space are expressed as follows,
\be
&&\delta\psi_\pm=\pm{i}\left(C_A+C_V\right)\psi_\pm,
\qquad\delta{A}_\pm=\mp\partial_\pm\left(C_A{\mp}C_V\right).
\ee
By requiring the equality between the BRST transformation in EPS and the
one in configuration space, we get the following relation
\be
{\cal C}^{\pm}=\mp\left(C_A\pm{C}_V\right).
\ee
The effective action Eq.(\ref{eff}) contains many non propagating
fields, i.e. $\pi^A_\pm$, ${\cal P}^\pm_\pi$, $\bar{{\cal P}}_\pm$,
$\bar{{\cal P}}^\pi_\pm$, ${N}^\pm$,${N}^\pm_\pi$, $\pi^A_\pm$ and
$\pi_\theta$, which can be eliminated by virtue of the equations of motion
 Eq.(\ref{eqmotion}). After eliminating these variables from the master
action, We arrive at the following BRST invariant effective action
in the most general class of gauges.
\be
\begin{array}{r c l@{}l}
S_{eff}
 &=&\ds\int dxdt:\biggl[&
    \psi^{\dag}_{+}\left({i}\partial_{-}+A_{-}\right)\psi_{+}
     +\psi^{\dag}_{-}\left({i}\partial_++A_+\right)\psi_{-}\\
 &&&-\ds\frac{1}{2}\partial_+\theta\partial_-\theta
     +k\left(A_-\partial_{+}\theta+A_{+}\partial_{-}\theta\right)
      -\frac{1}{4\pi}A_{-}A_{+}\qquad\\
 &&&-\ds\frac{1}{8\pi}\left(A_+^2+A_-^2\right)\qquad\\
 &&&-\ds\left(B_{+}\chi^{+}+\bar{{\cal C}}_{+}\delta\chi^{+}
     +B_{-}\chi^{-}+\bar{{\cal C}}_{-}\delta\chi^{-}\right)\,\biggr]: .\\
\end{array}\label{pg}
\ee
The effective action contains two types of counter term.
One is covariant type, which is constructed by the gauge fields and
BF fields. The other is non-covariant type, which is constructed by
only the gauge fields. The origin of appearance of non-covariant term
is related to the fact, that the manifest two dimensional covariance
is violated in the class of regularization schemes as we prove in
Appendix B. In order to recover the two dimensional covariance, we need
 an appropriately chosen noncovariant counter term.
It is nothing but this counter term.
The covariant counter term corresponds to one in the
Thirring model coupled to the gauge fields\cite{YW}.

In this section, we investigated the model as anomalous gauge
theory by applying the generalized Hamiltonian formalism of
BF and BFV, and got BRST invariant effective action.
This enables us to formulate the theory in most general class of gauges.

%%%%%%%%%%%%%%%%%%%%%%%%%%%%%%%%%%%%%%%%%%%%%%%%%%%%%%%%%%%%%%%%%%%%%%%%%%%%%%%

\section{Explicit gauge fixed effective action }
\subsection{conformal gauge}
\setcounter{equation}{0}
In the previous section we have formulated the BRST invariant
effective action. So far our argument did not rely on particular
gauge condition; the effective action can be applicable to any gauge fixing.
In this section, we will investigate the explicit gauge fixed theory.

First, we will consider the gauge fixed theory under ${A}_\pm=0$.
As the result, we show that the physical
excitation mode consists of only a null state as asserted in \cite{IM}.

\vspace{0.4cm}

We have recovered the local $U(1)_V\times U(1)_A$ symmetries
employing the BF formalism at quantum level.
Therefore we can choose the following gauge fixing condition
\begin{equation}
\chi_{\pm}=A_{\pm}.
\end{equation}
The gauge current of $A_\pm$ is given by
$J_\pm\equiv{j}_\pm+k\partial_\pm\theta$ in this gauge.

Integrating out the multipliers $B_\pm$ and eliminating non propagating
variables by the equations of motion, we can reduce the
effective action to the following expression
\be
S_{eff}=\int dxdt \left[\psi^{\dag}_+{i}\partial_{-}\psi_{+}+
\psi^{\dag}_-{i}\partial_{+}\psi_{-}-\frac{1}{2}\partial_{+}
\theta\partial_{-}\theta+\bar{{\cal C}}_+\partial_-{\cal C}^+
+\bar{{\cal C}}_{-}\partial_{+}{\cal C}^{-}\right].
\label{eac}
\ee
BRST charge is given by
\be
\begin{array}{c}
Q=Q_++Q_- , \quad
{\rm where}~{Q}_{\pm}=\ds\int dx{\cal C}^{\pm}J_\pm ,\\
\left[Q_+,Q_-\right]=0,\quad Q^2_\pm=0.\\
\end{array}
\ee
BRST invariant stress tensor $T_{\pm\pm}$ is constructed by the
Noether's theorem and algebra of stress tensor is $c=0$
Virasoro algebra.
Moreover, $T_{\pm\pm}$ is written in the BRST trivial form
\be
\begin{array}{rcl}
T_{\pm\pm}\left(x,t\right)
 =\left[Q_{\pm},X_\pm\left(x,t\right)\right],
\end{array}\label{exactform}
\ee
where ${X}_\pm\left(y,t\right)
 =-{i}/{4\pi}\bar{{\cal C}}_\pm\left(\psi^{\dag}_\pm\psi_\pm
     -k\partial_\pm\theta\right)$.

The fact in Eq.(\ref{exactform}) means that there is no physical
excitation state of the system.
\cite{IM}

\vspace{0.5cm}

%%%%%%%%%%%%%%%%%%%%%%%%%%%%%%%%%%%%%%%%%%%%%%%%%%%%%%%%%%%%%%%%%%%%%%%%%%%%%%%

\subsection{light-cone gauge}
In this subsection, we will show that a degree of freedom of
gauge field becomes a dynamical variable in the light-cone gauge
because of gauge anomaly at quantum level.

\vspace{0.25cm}

In the gauge fixing condition
\begin{eqnarray}
\chi^-=A_-,\quad\chi^+=\frac{\theta}{k},
\end{eqnarray}
the gauge current of $A_-$ is given by
$J_+=j_+-k^2A_+$.
The effective action is expressed by
\be
S_{eff}=\int{d^2}x\left[\psi^{\dag}_+{i}\partial_-\psi_{+}
+\psi^{\dag}_-{i}\partial_{+}\psi_{-}+A_+{j}_{-}-\frac{1}{8\pi}A^2_{+}-
\bar{{\cal C}}_-\partial_-C_A\right],
\ee
and the off-shell nilpotent BRST charge is given by
\be
Q&=&-\int{d}xC_AJ_+.
\ee
$Q$ is constructed by the fields, and depends only on $x^+$.
By using the equation of motion $\partial_-\bar{{\cal C}}_-=0$
and the BRST transformation for $\bar{{\cal C}}_-$,
the equation of motion for the gauge field $A_+$ is given by
\be
\partial_{-}A_+=0.\label{ce}
\ee
The Eq.(\ref{ce}) corresponds to the curvature equation in two
dimensional gravity.

The commutator for $A_+$ is given by
\be
\left[A_+\left(x\right),A_+\left(y\right)\right]=-8i\pi\partial_x
\delta\left(x-y\right). \label{etca}
\ee
as showed in Appendix D.

In this gauge, the stress tensor based on
the Noether's theorem does not generate the translation for fields.
However we can construct the true stress tensor as follows,
which generates the translation for the fields,
\be
\tilde{T}_{++}= :
\frac{1}{2}\left(\psi^{\dag}_+\partial_+\psi_+-\partial_+\psi^{\dag}_+
\psi_+\right)-\bar{{\cal C}}_-\partial_+C_A+\frac{1}{16\pi^2}A_+^2 : .
\ee
Here we have added the contribution of $A_+$,
\begin{eqnarray}
\frac{1}{16\pi^2}:A_+^2: \label{aterm}
\end{eqnarray}
into the stress tensor based on the Noether's theorem by hand.
$\tilde{T}_{++}$ satisfies the Virasoro algebra with $c=0$ and
it is expressed by the following BRST trivial form.
\be
\tilde{T}_{++}\left(x\right)=i\pi\left[Q,X\left(x\right)\right],
\ee
 where $X\left(x\right)=\bar{{\cal C}}_-\left(j_++1{/}4\pi{A}_+\right)$.

The stress tensor based on the Noether's theorem was not true stress tensor at
quantum level, because it did not generate any
translation for fields.
So we modified the stress tensor in analogy to Sugawara construction.
One can interpret Eq.(\ref{aterm}) as the Sugawara form for the gauge
field $A_+$ which satisfies the Kac-Moody algebra as Eq.(\ref{etca}).
Moreover, $\tilde{T}_{++}$ satisfies the Virasoro algebra with $c=0$, and
it generates the translation for fields at quantum level.
Hence, $\tilde{T}_{++}$  is regarded as the true BRST invariant stress tensor.
As the result, we recognize that the gauge field $A_+$
is a dynamical variable at quantum level.

%%%%%%%%%%%%%%%%%%%%%%%%%%%%%%%%%%%%%%%%%%%%%%%%%%%%%%%%%%%%%%%%%%%%%%%%%%%%%%%

\section{Conclusion}
\setcounter{equation}{0}
In this paper,
we investigated the massless Schwinger model without
kinetic terms of gauge field as an anomalous gauge theory by applying
the generalized Hamiltonian formalism of the BF and the BFV.
The BFV formalism  enables us to formulate the theory in most general
class of gauge fixing conditions.

Here, we describe the difference between our quantization and the
quantization in \cite{IM}.
In any anomalous gauge theory, the introduction of the Wess-Zumino
scalar is independent of the freedom of the gauge fields.
The BF field which has be introduced in section 2 corresponds to the
Wess-Zumino scalar in the conformal gauge.
Owing to the extra degree of freedom, the symmetry violated by anomaly
recovered even at quantum level.
Therefore we were able to choose the conformal gauge fixing condition
$A_\pm=0$
in this system.
In the conformal gauge, the physical state is only null state as asserted in
\cite{IM}.
However the property of the author's gauge fixing condition
is completely different from our one.
The authors parameterize $A_+$ in term of $U(1)$ group element $\phi$
which behaves as the Wess-Zumino scalar finally
\be
A_+=-\partial_+\phi.\nonumber
\ee
However, in our formulation, the gauge field $A_+$ is
completely independent of the
Wess-Zumino scalar $\theta$.
The behavior of the gauge field $A_+$ as the dynamical
variable is derived based on the equation of motion for the anti-ghost.

Our assertion is as follows.
If one consider $g_{++}$ instead of $A_+$, one can understand that this
model has the non-critical
string like property as follows.
In the Light-cone gauge fixed two dimensional gravity, the gravity field
$g_{++}$ becomes a dynamical variable because of the conformal anomaly,
though all degree of freedom of gravity field are decoupled due to three
local symmetries, Weyl symmetry and reparametrization invariance at
classical level.
On the other hand, in the Schwinger model without the kinetic term of
gauge field, the
gauge field $A_+$ becomes a dynamical variable because of the gauge
anomaly, though all degree of freedom of gauge field are decoupled due
to two local symmetries, $U(1)_V\times U(1)_A$.
Therefore we expect that the study of the gauge theory without kinetic
term may be useful to understand the non-critical string theory.

\vspace{0.4cm}

{\bf Acknowledgments}

We thank T.Fujiwara and Y.Igarashi for illuminating discussions,
N.Sakai for helpful discussions, for reading the manuscript.
We are also grateful to C.Itoi for explaining their paper and his comment,
S.Ding for reading the manuscript.

%%%%%%%%%%%%%%%%%%%%%%%%%%%%%%%%%%%%%%%%%%%%%%%%%%%%%%%%%%%%%%%%%%%%%%%%%%%%%%%

%\newpage
\vspace{0.4cm}
\appendix
\vspace{0.4cm}
\noindent
{\Large{\bf Appendix}}\\
\vspace{-1.0cm}
\setcounter{equation}{0}
\setcounter{footnote}{0}
\section{Alternative approach to BRST anomalies}

General gauge independent cohomological discussion
could be used for searching the anomalies \cite{FIK}.
In this Appendix, we review the general analysis of the BRST anomalies,
and apply to this model.

In general, BRST charge $Q$ and total Hamiltonian $H_T$ obey classical
gauge algebra.
BRST charge is constructed under the nilpotent condition,
\begin{eqnarray}
\left\{ Q,Q\right\} &=& 0,\label{QQ}
\end{eqnarray}
which means that the constraint which generate classical symmetry is
first class constraint.The consistency condition means that
BRST charge is time independent, or Hamiltonian is BRST invariant,
\begin{eqnarray}
\left\{ H_T, Q\right\} &=& 0.\label{Qdot}
\end{eqnarray}

At the quantum level, super-Poisson brackets must be replaced with
super-commutator s and operator must be regularized by the appropriate
prescription(e.g. normal ordering). But we do not suppose specific
regularization  prescription in this Appendix; so we can discuss the BRST
anomalies in the regularization independent way. If there are anomalies
in the system, Eq.(\ref{QQ}) and Eq.(\ref{Qdot}) are broken at quantum
level. The anomalous terms may be expanded in $\hbar$ as
\be
\begin{array}{rcl}
 \left[ Q,Q\right]&\equiv&i\hbar\W + O(\hbar^3 ), \\
 \left[ Q,H_T\right]&\equiv& i\hbar\G + O(\hbar^3 ).\\
\end{array}
\ee
We assume super-commutator for the operator must obey
the commutation law
\be
\bigl[ A, B\bigr] = -(-)^{\e(A)\e(B)}\bigl[ B, A\bigr],
\ee
to which the distribution law
\be
\bigl[ A, B+C \bigr] = \bigl[ A,B\bigr] + \big[ A, C\bigr],
\ee
and the super-Jacobi identity
\be
(-)^{\e (A)\e (B)}\left[ A,\left[ B,C\right]\right]
 +(-)^{\e (B)\e (C)}\left[ B,\left[ C,A\right]\right]
   +(-)^{\e (C)\e (A)}\left[ C,\left[ A,B\right]\right] =0
\ee
hold. By these identity, we easily find
\be
\begin{array}{c}
 \left[ Q,\left[ Q,Q\right]\right]\quad = \quad 0,\\
 2\left[ Q,\left[ Q,H_T\right]\right]
   +\left[ H_T,\left[ Q,Q\right]\right]\quad =\quad 0.\\
\end{array}
\ee
To the lowest $\hbar^2$ oder, super-Jacobi identities of $Q$ and $H_T$
can be truncated into the two conditions,
\be
\d\W &=& 0,\label{wzcc}\\
\d\G &=& - {\dot \W},\label{de}
\ee
where $\d$ is classical BRST transformation. Eq.(\ref{wzcc}) is called
Wess-Zumino consistency condition\cite{WZ}, and Eq.(\ref{de}) is called descent
equation\cite{DES}. We will find $\W$ in the exhaustive fashion, and construct
the most general solution of Wess-Zumino consistency condition in
regularization independent way.

If $\W$ is a solution of (\ref{wzcc}), then ${\tilde \W }\equiv \W +\d\X $
is a solution too. If we redefine the BRST charge as
\be
{\tilde Q}\equiv Q - {\hbar \over {\, 2\,}}\X.
\ee
The consistency condition is
\be
\left[ {\tilde Q},{\tilde Q}\right]\equiv i\hbar^2\W
\ee
We are interested in $\W $ which can not be expressed
by $\X$. The $\W$ is true anomaly.

We will search local form of anomalies,
\be
\W =\int dx \w.
\ee
In this case Putting $ C \equiv dim {\cal C}^{\pm}$, we find
\be
\begin{array}{c}
 dim {\cal C}^{\pm}_{\p} =  C+1 ,\quad
   dim {\overline {\cal P}}_{\pm}=1-C ,
     \quad dim {\overline {\cal P}}_{\pm}^{\p}=-C.
\end{array}
\ee
So, $dim Q = C$ , $gh Q = 1$ and $dim\W = 2C$ , $gh\W = 2$.
If $\w$ is a solution of Wess-Zumino consistency condition Eq.(\ref{wzcc}),
then
\be
\tilde{\w}\equiv\w +\d h_1 +\dl_1h_2
\ee
is a solution too. We are interested in $\w$ which can not be expressed
by $\d h_1$ , $\dl h_2$. The $\w$ is true anomaly. Here $dim\w
=2C+1$,$gh\w =2$ ,
\be
\begin{tabular}{|c|c|c|}
\hline
&$h_1$&$h_2$\\
\hline
dim&$C+1$&$2C$\\
\hline
gh\#&1&2\\
\hline
\end{tabular}
\ee
Using information of geometrization in Eq.(\ref{geom}) and BRST
transformation in Eq.(\ref{brsttr}), we get following table
\be
\begin{array}{c}
\begin{tabular}{l c@{~}c@{~}c c@{~}c@{~}c c@{~}c@{~}c }
     &$(A_{\pm}$&,&$\P_{\pm}^{A})$&$(\v_{\pm}$&,&$i\v_{\pm}^{\dagger})$\\
  dim&1&&0&1$\slash$2&&1$\slash$2\\
 gh\#&0&&0&0&&0\\
     &$({\cal C}^{\pm}$&,&$\overline{\cal P}_{\pm})$
     &$({\cal P}^{\pm}$&,&$\overline{\cal C}^{\pm})$
     &$(N^{\pm}$&,&$B_{\pm})$\\
  dim&$C$&&$1-C$&$C+1$&&$-C$&1&&0\\
 gh\#&1&&$-1$&1&&$-1$&0&&0\\
     &$({\cal C}^{\pm}_{\p}$&,&$\overline{\cal P}_{\pm}^{\p})$
     &$({\cal P}^{\pm}_{\p}$&,&$\overline{\cal C}^{\pm}_{\p})$
     &$(N^{\pm}_{\p}$&,&$B_{\pm}^{\p})$\\
  dim&$C+1$&&$-C$&$C+2$&&$-C-1$&2&&$-1$\\
 gh\#&1&&$-1$&1&&$-1$&0&&0\\
\end{tabular}
\end{array}
\ee
We divide EPS into two sectors
\be
\begin{array}{rcl}
{\it S_1}&:&
    {\rm consisting\, of\quad }\left( \v_{\pm}^{\dagger},\v_{\pm}\right)\quad
     \left( {\cal C}^{\pm},{\overline {\cal P}}_{\pm}\right) \\
{\it S_2}&:&
       {\rm consisting\, of\, all\, the\, other\, fields\,}\\
\end{array}\nn
\ee
The each sector is orthogonal to other sector. In other words, $\d$
operation of the each sector satisfies
\be
\d_1^2 = \d_2^2 = 0 ,\qquad \d_1\d_2+\d_2\d_1 = 0,
\ee
where $\d = \d_1 + \d_2$. The $S_2$ sector is BRST trivial, and there is
no non-trivial solution.
And we assume that the $\w$ preserve the invariance under the following
global discrete symmetry\footnote{If we take care of the result of
geometrization , $\dl_{\mp}{\cal C}^{\pm}=-{\cal C}^{\mp}_{\p}$ , then
$\dl_x\rightarrow -\dl_x$ is imposed },
\be
\pm\rightarrow\mp ,\quad\dl_x\rightarrow -\dl_x .
\ee
Therefore the true anomaly $\w$ is given by
\be
\w = \k \left( {\cal C}^+\dl_1{\cal C}^+
               -{\cal C}^-\dl_1{\cal C}^-\right) .
\ee
We find that $\w$ is not represented by $\d h_1$ and $\dl_1h_2$.
And we will find $\G = \ds\int dx\g$,
\be
\g = -4\k \left( N^+\dl_1{\cal C}^+-N^-\dl_1{\cal C}^-\right) .
\ee
This local form of $\G$ satisfies descent equation Eq.(\ref{de}).

%%%%%%%%%%%%%%%%%%%%%%%%%%%%%%%%%%%%%%%%%%%%%%%%%%%%%%%%%%%%%%%%%%%%%%%%%%%%%%%
\section{The definition of the normal ordering}
 The overall coefficient $\kappa$ in Eq.(A.19) or Eq.(A.20),
however, remains undetermined in the algebraic method in Appendix A.
To fix $\kappa$, we must define operator products by some ordering
prescription, and then examine the nilpotency of the BRST charge.
For the purpose to define operator ordering,
we decompose operators into parts by
\be
&&{F}^{(\pm)}\left(x\right)=\int dy\delta^{(\mp)}\left(x-y\right)
{F}\left(y\right)~;~{\rm for}~{F}=\psi_+,\psi_+^{\dag},\bar{{\cal C}}_+,
{\cal C}_+,\pi_{\theta}+\theta{'}\nonumber\\
&&{F}^{(\pm)}\left(x\right)=\int dy\delta^{(\pm)}\left(x-y\right){F}
\left(y\right)~;~{\rm for}~{F}=\psi_-,\psi_-^{\dag},\bar{{\cal C}}_-,
{\cal C}_-,\pi_{\theta}-\theta{'}\nonumber
\ee
where $\delta^{\left(\pm\right)}\left(x\right)=
\pm{i}/2\pi\left(x\pm{i}\epsilon\right)$. The $F^{(+)}$ and $F^{(-)}$
thus defined reduce, respectively, to positive- and negative-frequency part
in the conformal gauge. By putting $F^{(+)}$'s to the right of $F^{(-)}$'s,
we define operator ordering. We thus obtain
\be
\k^2=1{/}{4\pi}\nonumber
\ee
for the anomaly coefficient.

%%%%%%%%%%%%%%%%%%%%%%%%%%%%%%%%%%%%%%%%%%%%%%%%%%%%%%%%%%%%%%%%%%%%%%%%%%%%%%%
\section{The derivation of the commutator for $A_+$}
\setcounter{equation}{0}

The BRST transformation of $A_+$, which has been read off from
Eq.(\ref{brsttr}) and the variation for the anti-ghost is given by
\be
\delta{A}_+=\partial_+C_A.\label{delA}
\ee
On the other hand, $\delta{A}_+$ is defined by
\be
\delta{A}_+\left(x^+\right)=-i\bigl[A_+\left(x^+\right),Q_A\bigr]
\quad{\rm where}\quad{Q}_A=\frac{1}{4\pi}\int{d}x^+C_AA_+.
\ee
Comparing this with $\delta{A}_+$ given in Eq.(\ref{delA}),
we obtain the commutation relation Eq.(3.10).

%%%%%%%%%%%%%%%%%%%%%%%%%%%%%%%%%%%%%%%%%%%%%%%%%%%%%%%%%%%%%%%%%%%%%%%%%%%%%%%
\section{Quantum BRST algebra in conformal gauge}

In this subsection, we will discuss the quantum BRST algebra of the system.
It might be another approach to construct quantum topological field
theory. \\
\vspace{10 pt}
In the conformal gauge, the BRST charge is given by
\be
Q = Q_+ +Q_-, \quad {\rm where } \quad
Q_{\pm}=\int dx{\cal C}^{\pm}(j_{\pm}+k\dl_{\pm}\c).
\ee
We define anti-BRST charge \cite{abrst} using uncertain sign of the level $k$.
\be
\overline{Q} = \overline{Q}_++\overline{Q}_-, \quad {\rm where } \quad
\overline{Q}_{\pm}=\int dx{\cal C}^{\pm}(j_{\pm}-k\dl_{\pm}\c).
\ee
Ghost number charge is
\be
Q_c=\int dx\left[\overline{\cal C}_{+}{\cal C}^+
                 +\overline{\cal C}_{-}{\cal C}^-\right].
\ee
The quantum extended BRST algebra is
\be
\begin{array}{c}
  \begin{array}{c@{~=~}c@{\qquad}c@{~=~}c}
    \bigl[~Q, Q ~\bigr] & 0,&\bigl[ ~\overline{Q},\overline{Q} ~\bigr] & 0,\\
    \bigl[~iQ_c, Q ~\bigr] & +Q,
      &\bigl[ ~iQ_c,\overline{Q} ~\bigr] & -\overline{Q},\\
  \end{array}\\
  \begin{array}{rcl}
    \bigl[~Q_c , Q_c~\bigr] &=&  0,\\
    \bigl[~Q_{\pm},\overline{Q}_{\pm}~\bigr] &=& 4\p i \ds\int dx T_{\pm\pm}.\\
  \end{array}
\end{array}
\ee
The extended BRST algebra and its representation has been well studied in
classical level \cite{ebrst}. The quantum extended BRST algebra
was studied in the flame of conformal field theory. This algebra
correspond to $c = 0$ trivial topological field theory discussed in
\cite{eguti}, and this model give  dynamical representation of $c = 0$
trivial topological field theory.

%%%%%%%%%%%%%%%%%%%%%%%%%%%%%%%%%%%%%%%%%%%%%%%%%%%%%%%%%%%%%%%%%%%%%%%%%%%%%%%

\end{document}